\documentclass[final,1p,times]{elsarticle}

\usepackage{epsfig} 
\usepackage{graphics}
\usepackage{xspace}
\usepackage{amsmath}
\usepackage{subfigure}
\usepackage{booktabs}
\usepackage{color}

\definecolor{Red}{rgb}{0.7,0.0,0.0}
\definecolor{Green}{rgb}{0.0,0.7,0.0}
\definecolor{Blue}{rgb}{0.0,0.0,0.7}

\usepackage{amssymb}

\journal{Physica A}

\begin{document}

\begin{frontmatter}



\title{Small World Picture of Worldwide Seismic Events}


\author[ifrj,on]{Douglas S. R. Ferreira}
\ead{douglas.ferreira@ifrj.edu.br}

\author[on,uerj]{Andr\'es R. R. Papa\corref{cor}}
\ead{papa@on.br}

\author[fit]{Ronaldo Menezes}
\ead{rmenezes@cs.fit.edu}

\cortext[cor]{Corresponding authors}

\address[ifrj]{Instituto Federal de Educa\c{c}\~ao, Ci\^encia e Tecnologia do Rio de Janeiro, Paracambi, RJ, Brazil}

\address[on]{Geophysics Department, Observat\'orio Nacional, Rio de Janeiro, RJ, Brazil}

\address[uerj]{Instituto de F\'isica, Universidade do Estado do Rio de Janeiro, Rio de Janeiro, RJ, Brazil}

\address[fit]{BioComplex Laboratory, Computer Sciences, Florida Institute
  of Technology, Melbourne, USA}

\begin{abstract}
The understanding of long-distance relations between seismic
activities has for long been of interest to seismologists and
geologists. In this paper we have used data from the world-wide
earthquake catalog for the period between 1972 and 2011 to generate a
network of sites around the world for earthquakes with magnitude $m
\geq 4.5$ in the Richter scale. After the network construction, we
have analyzed the results under two viewpoints. Firstly, in contrast
to previous works, which have considered just small areas, we showed
that the best fitting for networks of seismic events is not a pure
power law, but a power law with exponential cutoff; we also have found
that the global network presents small-world properties. Secondly, we
have found that the time intervals between successive earthquakes have
a cumulative probability distribution well fitted by nontraditional
functional forms. The implications of our results are significant
because they seem to indicate that seisms around the world are not
independent. In this paper we provide evidence to support this argument.
\end{abstract}


\end{frontmatter}





\section{Introduction}

The general belief in seismic theory is that the relationship between
events that are located far apart is hard to be
understood/demonstrated. However we live today in a world where data
is being collected on most aspects of our lives and better yet,
computer power is cheaply available for analyzing such data. When we
apply the computer power to the data we open a series of possibilities
to look for patterns in the data. The work on seismic data analysis is
no different; we have now large collections of millions of seismic
events from around the world each of which deserving deeper analysis. In this
paper we have found some evidence that point to small-world
characteristics in the existing data on seismic events. An event in a
particular geographical site appears to be related to many other sites
around the world and not only to other events at nearby sites.


The ability to find useful information from data is not new and is commonly known as Data Mining. However, since the work from Barab\'asi and Albert \cite{barabasi99} researchers have turned their attention not to mining the data itself but rather organizing the data in a network which captures relationships between pieces of data and only then mining the network structure and hence the relations between pieces of data. The network may review patterns that could not be observed from mining the raw pieces of data. The use of networks as a framework for the understanding of natural phenomena is nowadays called {\it Network Science}.

In the last few years, some analysis related to seismic phenomena demonstrated that earthquakes display features explained from the perspective of non-extensive statistical mechanics \citep{abe03,abe05,darooneh08,darooneh10,vallianatos2013evidence}. These networks display complex features that can be better understood statistically using the Tsallis entropy \citep{tsallis1988}. Through the analysis of distances and time intervals between successive earthquakes using non-extensive statistical mechanics, the authors have found that two successive earthquakes are indivisibly correlated, no matter how much spatially far they are from each other \citep{abe06}.

In line with the successive earthquake model mentioned above, recent
studies \citep{abe04_a,abe04_b} have applied concepts of complex
networks to study the relationship between seismic events. In these
studies, networks of geographical sites are constructed by choosing a
region of the world (e.g. Iran, California) and its respective
earthquake catalog. The region is then divided into small cubic cells,
where a cell will become a node of the network if an earthquake
occurred therein. Two different cells will be connected by a directed
edge when two successive earthquakes occurred in these respective
cells. If two earthquakes occur in the same cell we have a loop, i.e.,
the cell is connected to itself. Fig.~\ref{fig:network} depicts a toy
example of a network being formed. This method of describing the
complexity of seismic phenomena has found that, at least for some
regions, the common features of complex networks (e.g. scale-free,
small-world) are present. However, in spite of the importance of the
results, they are somewhat expected, since it makes sense for areas located geographically near to each other, to be correlated. 

In this paper we have used data from the world-wide earthquake catalog
for the period between 1972 and 2011, to generate a network of sites
around the world. Since only seismic events with $m \geq 4.5$ are
recorded for all locations around the world, we then consider them
{\it significant events} and used this set in our analysis (all seisms
with $4.5$ or more on the Richter scale. The results were analyzed
under two viewpoints. The first, under the perspective of complex
networks theories, and the second using non-extensive statistical
mechanics.

\section{Theoretical Background}

\subsection{Complex Networks Features}
\label{complex_net}

Scale-free networks are defined as those in which the degree distribution of nodes (or vertices) follows a power law, that is, the probability that a network will have nodes of degree $k$, denoted by $P(k)$ is given by
\begin{equation}
\label{power_law}
P(k) \sim k^{-\gamma},
\end{equation}
where $\gamma$ is a positive constant.

Eq.\,\ref{power_law} states that scale-free networks have a very small number of highly-connected nodes (called hubs) and a large number of nodes with low connectivity. These networks exist in contrast with general random networks  with a very large number of nodes in which the probability distribution follows a Poisson distribution.
\begin{equation}
\label{random-network}
P(k) = \binom{N}{k} p^k(1-p)^{N-k} \simeq\frac{\langle k \rangle^k e^{-\langle k \rangle}}{k!},
\end{equation}
where $N$ is the number of nodes in the network and each node has an average of $\langle k \rangle$ connections. In Eq.\,\ref{random-network}, $p$ represents the probability of an edge to be present in the network and can be shown to be approximately $\langle k \rangle/N$. Random networks have nice properties but the truth of the matter is that most real networks are not random. 

The definition of a small-world network is yet to be formalized. One of the best approaches for defining small-world networks is based on the work of Watts \citep{watts1999networks} which states that in small-world networks, every node is ``close'' to every other node in the network. It is generally agreed that ``close'' refers to the average path length in the network, $\ell$, having the same order of magnitude as the logarithm of the number of nodes, i.e., 
\begin{equation}
\label{avg_pathlength}
\ell \sim \ln N. 
\end{equation}


In addition, and what makes small-world networks even more interesting is the fact that they have a high degree of clustering representing a transitivity in the relation of nodes; if a node $i$ has two connections, the theory argues that the two connections are also likely to ``know" each other. More formally, the clustering coefficient, $C_i$, of that node is given by:
\begin{equation}
C_{i} = \frac{ \triangle(i) }{ \triangle_{all} (i) }
\label{clust_coef}
\end{equation}
where, $\triangle (i)$ is the number of the directed triangles formed by $i$ with its neighbors and $\triangle_{all} (i)$ is the number of all possible triangles that $i$ could form with its neighbors; the clustering coefficient of the entire network, $C$, is just the average of all $C_i$ over the number of nodes in the network, $N$. In random networks the clustering coefficient can be estimated using the closed form 
\begin{equation}
  \label{eqn:crand}
  C_{\text{rand}}= {{\langle k \rangle} \over {N}},
\end{equation}
where $\langle k \rangle$ is the average degree in the random network.

Last, one needs to understand the relation of these two characteristic to the world of seismic events. If a network of seismic events contains hubs, one can argue that the distribution of earthquakes should also follow a power law. On the other hand, if the network of seismic events has small-world properties one can argue that there is some indication of long-range relations between far-apart earthquake sites.

\subsection{Summary on Non-extensive Statistical Mechanics}
\label{nonextensive}

Nonextensive statistical mechanics is a theory introduced to explains many physical systems where the traditional Boltzmann-Gibbs statistical mechanics does not seems to apply. This theory can explain a variety of complex systems with characteristics such as long-range interaction between its elements, long-range temporal memory, fractal evolution of phase space, and certain kinds of energy dissipation. In these cases, we use Tsallis entropy \citep{tsallis1988}, which is a generalization of the Boltzmann-Gibbs entropy (defined later in this paper). The Tsallis entropy is defined as: 
\begin{equation}
\label{tsallis_entropy}
S_q = K \frac{1 - \sum_{i=1}^{W}p_{i}^{q}}{q - 1},
\end{equation}
where $W$ is the total number of possible configurations, $q$ is the entropic index, $p_{i}$ are the associated probabilities and $K$ is a conventional positive constant. This entropy violates the additivity property, i.e. the entropy of the whole system can be greater or smaller than the sum of the entropies of its parts. In other words, if we have a system composed of two statistically independent subsystems $A$ and $B$,
\begin{equation}
\label{tsallis_additivity}
S_{q}(A+B) = S_{q}(A) + S_{q}(B) + \frac{(1 -q)}{K} S_{q}(A) S_{q}(B),
\end{equation}
where we can see that $q$ appears to characterize universality classes of nonadditivity.

Taking the limit $q \rightarrow 1$ in Eq. \ref{tsallis_entropy}, we get the standard Boltzmann-Gibbs entropy, $ S = -K \sum_{i=1}^{W} p_{i}\, \mathrm{ln}\,p_i $. It is also known that applying the maximum entropy principle to the Tsallis entropy, the probability distribution obtained has a $q$-exponential form \citep{abe03geometry}, $e_{q}(x)$, defined by,
\begin{equation}
\label{q_exp}
e_{q}(x) = \left\{ 
  \begin{array}{l l}
    [1 + (1 - q)x]^{1/(1 - q)}	& \quad \text{if} \quad [1 + (1 - q)x] \geq 0\\
    0			 & 	\quad \text{if} \quad [1 + (1 - q)x] < 0
  \end{array} \right.
\end{equation}

The inverse function of the $q$-exponential is the $q$-logarithmic function,
\begin{equation}
\label{q_ln}
ln_{q}(x) = \frac{x^{1-q} - 1}{1 - q},
\end{equation}
In the limit $q \rightarrow 1$ Eqs. \ref{q_exp} and \ref{q_ln} get back the standard exponential and logarithmic functions, respectively.

\section{A Geographical Network from Seismic Events}
\label{construction}

The use of networks to understand phenomena associated with geographical locations has been used in many instances in science including diseases \citep{newman2002spread}, scientific collaborations \citep{divakarmurthy2011temporal,pan2012world}, and organ transplantation \citep{Venugopal2012} to mention just a few. Seismic activity is intrinsically linked to geography because today's instruments can pinpoint with great accuracy the location in the globe where each seismic event takes place.

\begin{figure}
\begin{center}
\includegraphics[width=0.6\columnwidth]{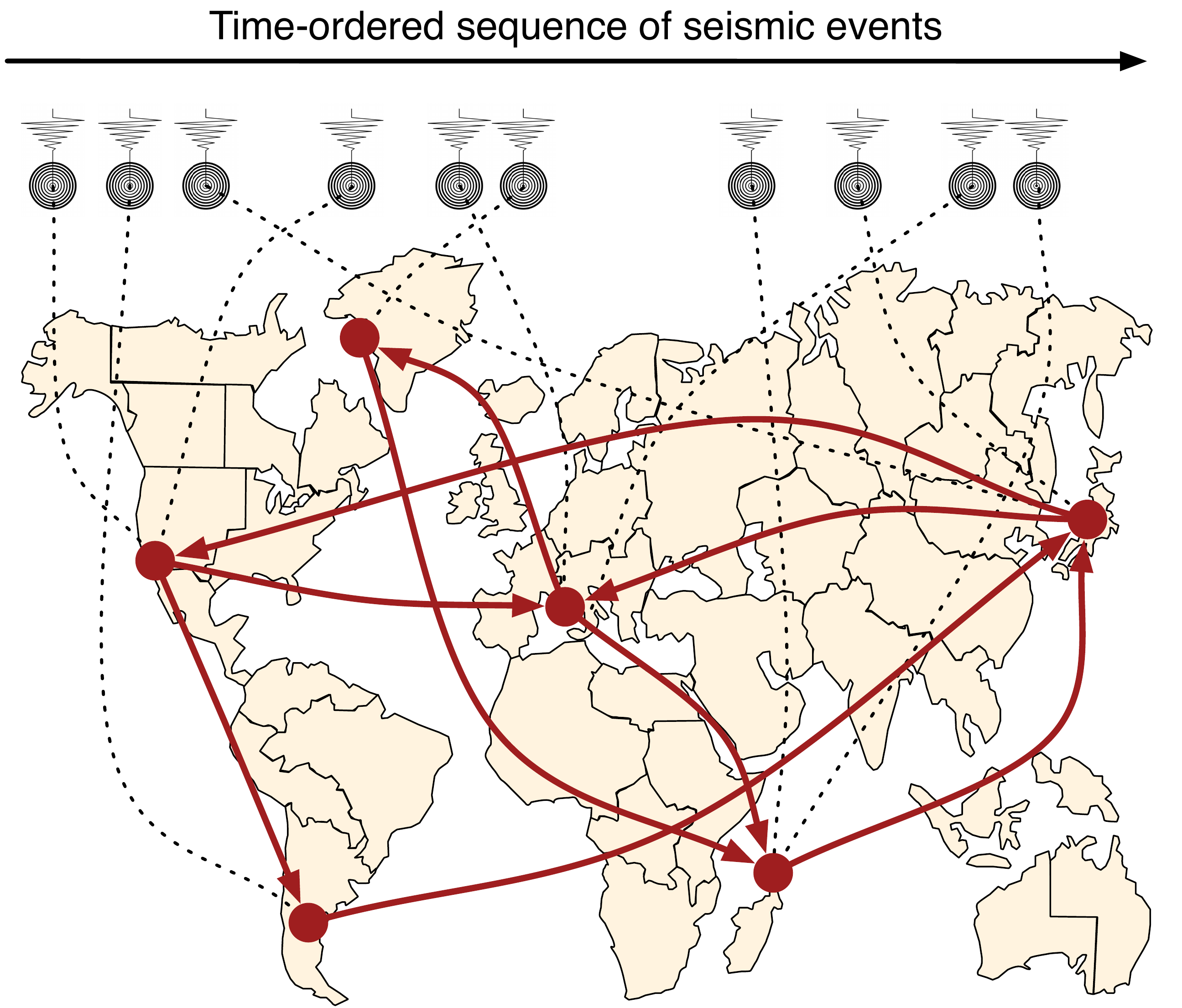}
\caption{A sketch of how the network of seismic events is created. At the top of the figure we see a sequence of time-ordered seismic events. Since each event has an epicenter $E$ with location $(\theta_E, \phi_E)$ we can use Equation \ref{S} to calculate which cell in the map to be used as a node in the network. The nodes are linked based on the sequence of events shown at the top of the picture.}
\label{fig:network}
\end{center}
\end{figure}

It is important to precisely locate the geographical location of a
seismic event but if we want to understand relations between events we
should concentrate on creating a network in which locations are linked
based on an acceptable criteria. In this paper, we use the same
procedure employed by Abe and Suzuki~\citep{abe04_a} in their studies
of earthquakes in specific regions of the world. The construction of
the network is as follows. We first have to decide on what should
represent the nodes. Obviously our first choice are the sites where
the earthquake took place. The problem of doing this is that an
earthquake epicenter is rarely located exactly in the same location
and given the accuracy of today's instruments we would have an
infinitely large number of possible sites. We decide instead to define
nodes representing a larger region of the world we here call a {\it
  cell}. A cell will become a node of the network if an earthquake has
its epicenter therein. The creation of edges follows a temporal order
of seismic events. For instance, if an earthquake occurs in a cell
$C_1$ and the next earthquake in a cell $C_2$, we assume a relation
between $C_1$ and $C_2$ and we represent the event by a directed edge
in the network. The process continues linking cells according to the
temporal order. Fig.~\ref{fig:network} depicts the process used to
create the network from seismic events. It is worth noting that if two
successive earthquakes occurs in the same cell, this node will be
connected to itself via a self-edge or a loop.

The degree of each node (the total number of its connections) is not affected by the direction of the network. The nature of the way the network is constructed means that for each node in the network, its in-degree is equal to its out-degree (the exceptions are only the first and last sites in the sequence of seismic events but for all practical purposes we can disregard this small difference). 

Although the use of temporal ordering of events is not new in our paper, there are two main differences between our study and others. First and most importantly, the region considered in our investigation is the entire globe, instead of just some specific geographical subarea of the globe; this is, to the best our knowledge, the first worldwide study of seismic events using networks and consequently the first one to investigate the possibility of long-range links between seismic events. Second, we have used a two dimensional model in which the depth dimension of the earthquake epicenter is not considered, since we are interested in looking for spatial connections between different regions around the world and besides 82\% of the earthquakes, in our dataset, have their hypocenters in a depth less than or equal to 100\,km.


Before we divide the globe into cells, we need to choose the size of such cells particularly because we are dealing with the entire globe; if the cells are too small we will not have any useful information in the network, if the cells are large we lose information due to the grouping of events into a single network node. There are no rules to define the sizes. Therefore we have taken three different sizes, the same sizes used in previous studies~\citep{abe04_a,abe10}, where the authors conducted studies about earthquake networks using data from California, Chile and Japan. The quadratic cells have, $\text{5\,km} \times \text{5\,km}$, $\text{10\,km} \times \text{10\,km}$ and $\text{20\,km} \times \text{20\,km}$. To set up cells around the globe, we have used the latitude and longitude coordinates of each epicenter in relation to the origin of the coordinates, i.e., where both latitude and longitude are equal to zero (we have chosen the referential at the origin for simplicity). So, if a seismic event occurs with epicenter $E$ with location $(\theta_E, \phi_E)$, where $\theta_E$ and $\phi_E$ are the values of latitude and longitude in radians of the epicenter, we are able to find the distances north--south and east--west between this point and the origin. These distances can be calculated, considering the spherical approximation for the Earth, by: 
\begin{equation}
\label{S}
\begin{array}{l}   
S^{ns}_E = R.\theta_E
\\
S^{ew}_E = R.\phi_E.\cos{\theta_E},
\end{array}
\end{equation}
where $S^{ns}_E$ and $S^{ew}_E$ are, the north-south and east-west distances for the earthquake $E$, respectively, and $R$ is the Earth radius, considered equal to $6.371\times 10^3$ km. With this computation we can identify the cell in the lattice for each event using the values of $S^{ns}_E$ and $S^{ew}_E$.

Note that the distances between different cells are irrelevant for the present part of our study. By now we are just interested in the connectivity of nodes. However, from the sequence there are important consequences to be obtained which we present below.

The seismic data used to build our network was taken from the Global Earthquake Catalog, provided by U.S Geological Survey (USGS), specifically at Advanced National Seismic System\footnote{http://quake.geo.berkeley.edu/anss}, which records events from the entire globe. The data spans all seismic events between the period from January 1, 1972 to December 31, 2011. This catalog has a limitation because it is not consistent in all regions of the world; it includes events of all magnitudes for the United States of America but only events with $m \geq 4.5$ (in the Richter scale) for the rest of the world (unless they received specific information that the event was felt or caused damage). Therefore, in order to obtain a more homogeneous distribution of data through the world, we have analyzed only events with $m \geq 4.5$. We have considered in our data the magnitudes Mb, ML, Ms and Mw, but we have excluded data that represent artificial seismic events (``quarry blasts'' and nuclear blasts). In the end, we were left with 185\,747 events, where 82\% of them happen near the surface of the world (depth~$\leq$~100\,km).

\section{Results}
\label{results}

Given the network build as described in Section \ref{construction}, we
have performed a few experiments to understand its
structure. Following the procedure depicted in Fig.~\ref{fig:network},
the 185\,747 seisms yielded three different networks depending on the
size used for the cells. Table~\ref{tab:networks} presents the sizes
of the three networks.

\begin{table}[ht]
  \caption{Three networks have been created from 185\,747 seismic events in our dataset. $N$ represents the number of nodes and $M$ represents the number of edges. Since we have a network constructed from consecutive seismic events, the number of edges, $M$, is always 1 less than the number of events.}
  \begin{center}
    \resizebox{0.4\columnwidth}{!}{%
      \tabcolsep=0.2cm
      \begin{tabular}{l c c}
        \hline
        Network 	&  $N$	&  $M$	\\
        \hline
        $\text{20\,km} \times \text{20\,km}$ &  65\,355 &  185\,746  \\
        $\text{10\,km} \times \text{10\,km}$ &  104\,516 & 185\,746  \\
        $\text{5\,km} \times \text{5\,km}$ &  144\,974 &  185\,746  \\
        \hline
      \end{tabular}%
    }
  \end{center}
  \label{tab:networks}
\end{table}

\subsection{Scale-Free Property of the Seismic Network}
\label{scale_free_cutoff}

It has been shown recently \citep{abe06,abe10} that seismic networks for specific regions of the globe (e.g. California) appear to have scale-free properties, or in other words that the construction of the network employs preferential attachment as described by \citep{barabasi99} insofar that a node added to the network has a higher probability to be connected to an existing node that already has a large number of connections. This is somewhat trivial to understand because active sites in the world will tend to appear in the temporal sequence of seismic events many times. The preferential attachment states that the probability $P$ that a new node $i$ will be linked to an existing node $j$, depends on the degree $\text{deg}(j)$ of the node $j$, that is, $P(i \rightarrow j) = \text{deg}(j) / \sum_{u}\text{deg}(u)$. This rule generates a scale-free behavior whose connectivity distribution follows a power-law with a negative exponent as shown in Eq.\,\ref{power_law}.

In \citep{abe06,abe10}, earthquake networks were built for some
specific regions (California, Chile and Japan), and their connectivity
distributions were found to follow power-laws. However, if we look
carefully to the connectivity distribution and plot its cumulative
probability, instead of its probability density, we can observe that
the power-law distributions that emerge from these network are
truncated. According with \citep{Amaral:Classes:2000}, there are at
least two classes of factors that may affect the preferential
attachment and consequently the scale-free degree distribution: the
aging of the nodes and the cost of adding links to the nodes (or the
limited capacity of a node). The aging effect means that even a highly
connected node may, eventually, stop receiving new links as normally
occurs in scientific collaboration networks \citep{newman01} where
scientists with time stop forming new collaborations maybe due to
retirement or because they are already satisfied with the number of
collaborators they have. The presence of an aging-like effect in our
work could be expected from the fact that relaxation times for
tectonics are much longer that the time interval under study; some
cells can stop of receiving new connections during a period of time
comparable to our own time window by a temporal quiet period due to a
transitory stress accumulation. The second factor that affects the
preferential attachment occurs when the number of possible links
attaching to a node is limited by physical factors or when this node
has, for any reason, a limited capacity to receive connections, like
in a network of world airports. We have not found a suitable parallel
to this factor in the case of earthquakes. These factors impose a
constraint to the preferential attachment and its power-law
distribution. When any of these factors is present, the distribution
is better represented with a power law with an exponential cutoff,
\begin{equation}
\label{power_law_exp}
P(k) \sim k^{-\alpha}e^{-k/k_c},
\end{equation}
where $\alpha$ and $k_c$ are constants.

\begin{figure}
  \begin{center}
    \subfigure[] {
    \includegraphics[width=0.6\columnwidth]{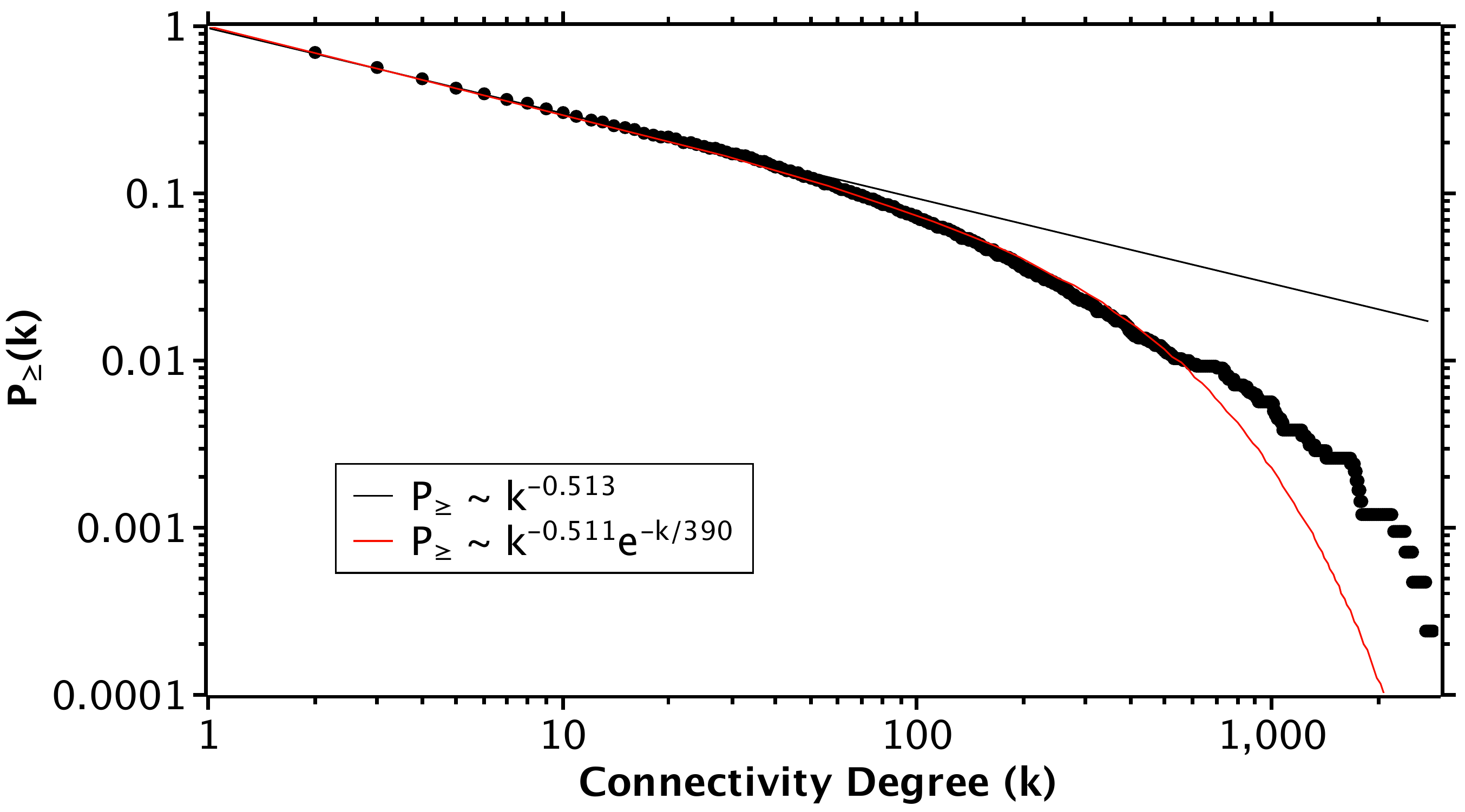}
      \label{k_california_all}}
    \subfigure[]  {
    \includegraphics[width=0.64\columnwidth]{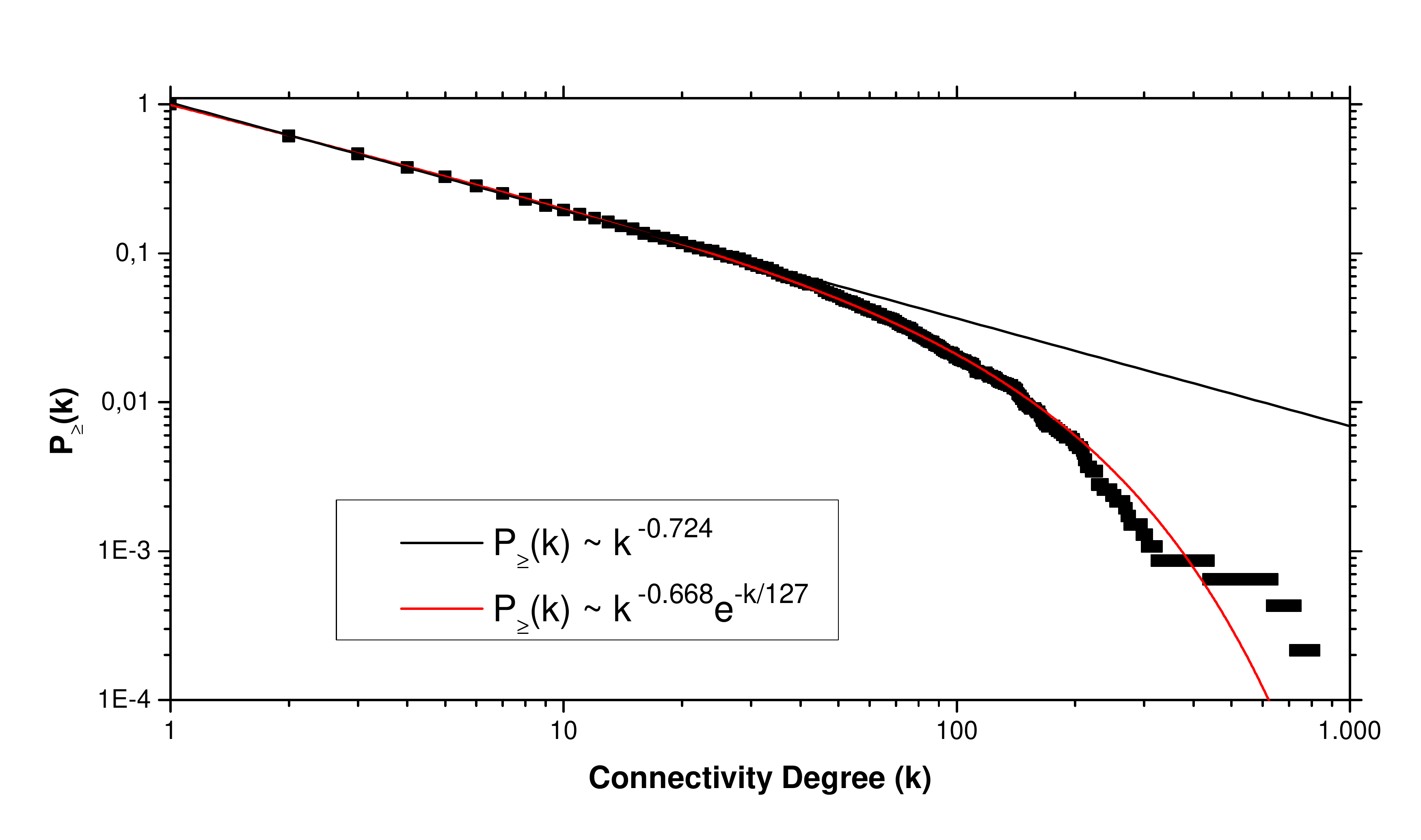}
      \label{k_california_larger25}}
    \caption{Cumulative probability distribution of connectivity for the earthquake network in California using cell size $\text{5\,km} \times \text{5\,km}$. The solid lines represent two possible fittings: a power-law (black) and a power-law with exponential cutoff (red). \subref{k_california_all} For the period between January 1, 2002 and December 31, 2011. The best fitting is for a power-law with exponential cutoff with $\alpha = -0.511 \pm 0.001$ and $k_{c}= -390 \pm 3$. There are 4\,187 nodes in this network. \subref{k_california_larger25} Considering just earthquakes with $m \geq 2.5$, for the period between January 1, 1972 and December 31, 2011. There are 4\,646 nodes in this network. The best fitting is for a power-law with exponential cutoff with $\alpha = -0.668 \pm 0.001$ and $k_{c}= -127 \pm 1$. }
    \label{k_california}
  \end{center}
\end{figure}

In Fig.~\ref{k_california_all}, we plot the cumulative probability distribution for the earthquake network built for the Southern California ($32^\circ \mathrm{N} - 37^\circ \mathrm{N}$ and $114^\circ \mathrm{W} - 122^\circ  \mathrm{W}$), using the data catalog provided by Advanced National Seismic System, where we considered all seisms with magnitude $m > 0$ for the period between January 1, 2002 and December 31, 2011. The total number of events are 147\,435. It is possible to observe in this plot that the data is better fitted to a power-law with exponential cutoff than a pure power law which is a good fit only for small values of $k$ with an exponent $\gamma - 1 = 0.513$, which is consistent with the value $\gamma = 1.5$ reported in \citep{abe06} for the probability density function. These results apply for a network built using cell sizes of $\text{5\,km} \times \text{5\,km}$. It is noteworthy that in a probability density plot, the cutoff does not seem to exist, because the fluctuations are higher than in a cumulative probability plot. Here we point out that for small magnitudes ($m<2.5$), the magnitude distribution does not follow the Gutember-Richter law, but a {\it{q-exponential}} distribution \cite{darooneh10}. Thereat, we also plot the cumulative probability distribution for the Southern California considering just earthquakes with $m \geq 2.5$, for the period between January 1, 1972 and December 31, 2011, which give us a total number of events equal 50\,847, as showed in Fig.~\ref{k_california_larger25}. It is interesting to note that in both cases we have a better fitting for a power-law with exponential cutoff than for a pure power law.

Before constructing the network for the entire world, we verified if the magnitude distribution of seismic events in our data has the expected behavior. The Gutenberg-Richter law gives the rate of occurrence of earthquakes with magnitude larger than or equal to $m$,
\begin{equation}
\label{mag}
F_\geq (m) = 10^{a - b.m},
\end{equation}
where, $F_{\geq}$ is the number of earthquakes with magnitude larger than or equal to $m$ and $a$ and $b$ are constants.

\begin{figure}
\begin{center}
\includegraphics[width=0.6\columnwidth]{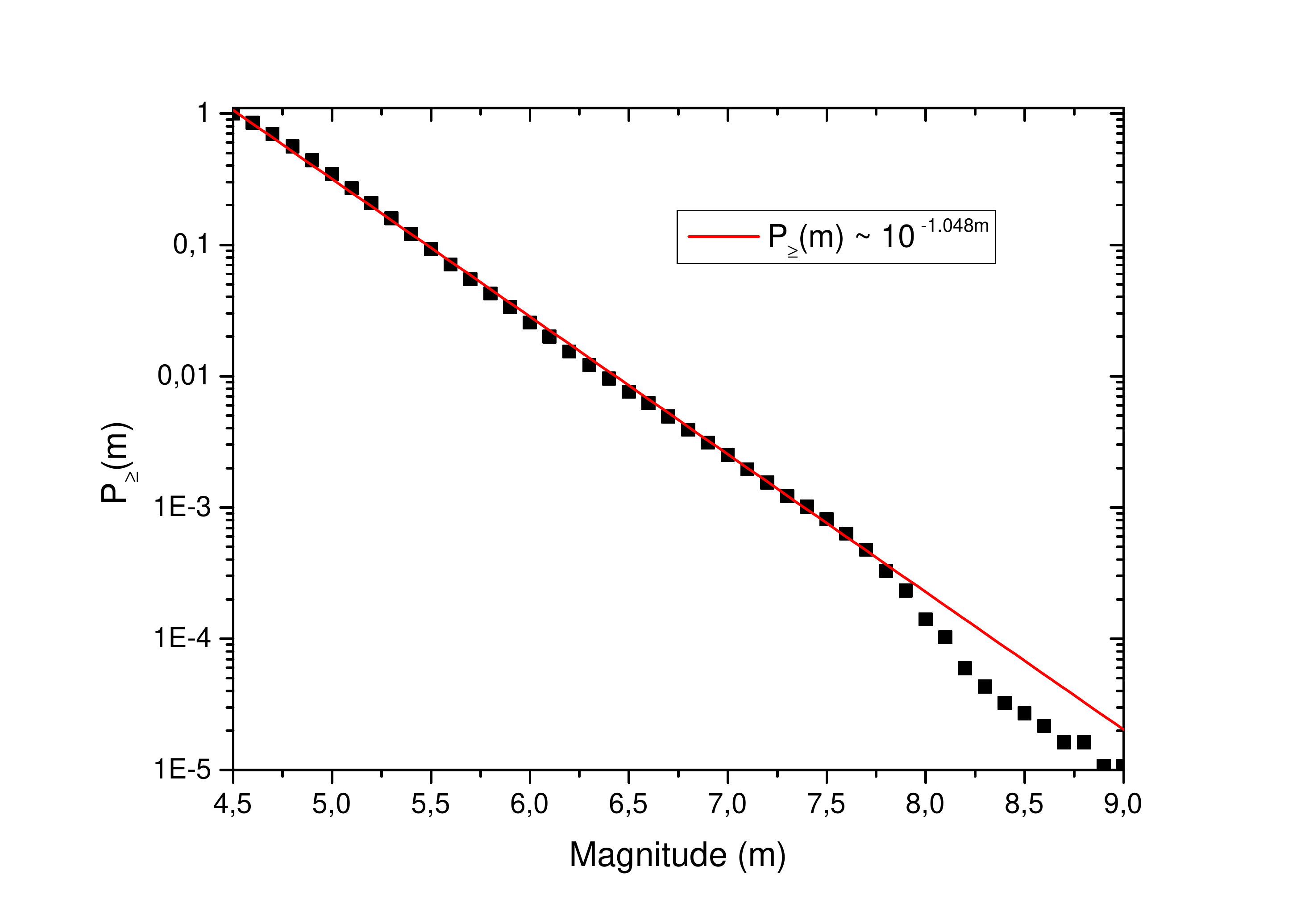}
\caption{Log-linear plot for cumulative probability distribution of the magnitude of earthquakes for the data used in this paper. The parameter $b$ from the Gutenberg-Richter law (GR), has the value $-1.048 \pm 0.002$, which was calculated using the method of maximum likelihood.}
\label{mag_dist}
\end{center}
\end{figure}

As described in \citep{darooneh10}, this equation presents problems only for small values of magnitude. Since for the globe we are using data with $m \geq 4.5$, it is expected that our magnitude data have a good agreement with the Gutenberg-Richter law. This agreement is shown in Fig.~\ref{mag_dist}, which gives the cumulative distribution of magnitudes and, using the maximum likelihood approach, we have found $b = -1.048 \pm 0.002$ (we have also calculated this {\it b-value} by the weighted least-square method, which gives $b= -1.139 \pm 0.011$).

\begin{figure}[h]
  \begin{center}
    \subfigure[] {
    \includegraphics[width=0.6\columnwidth]{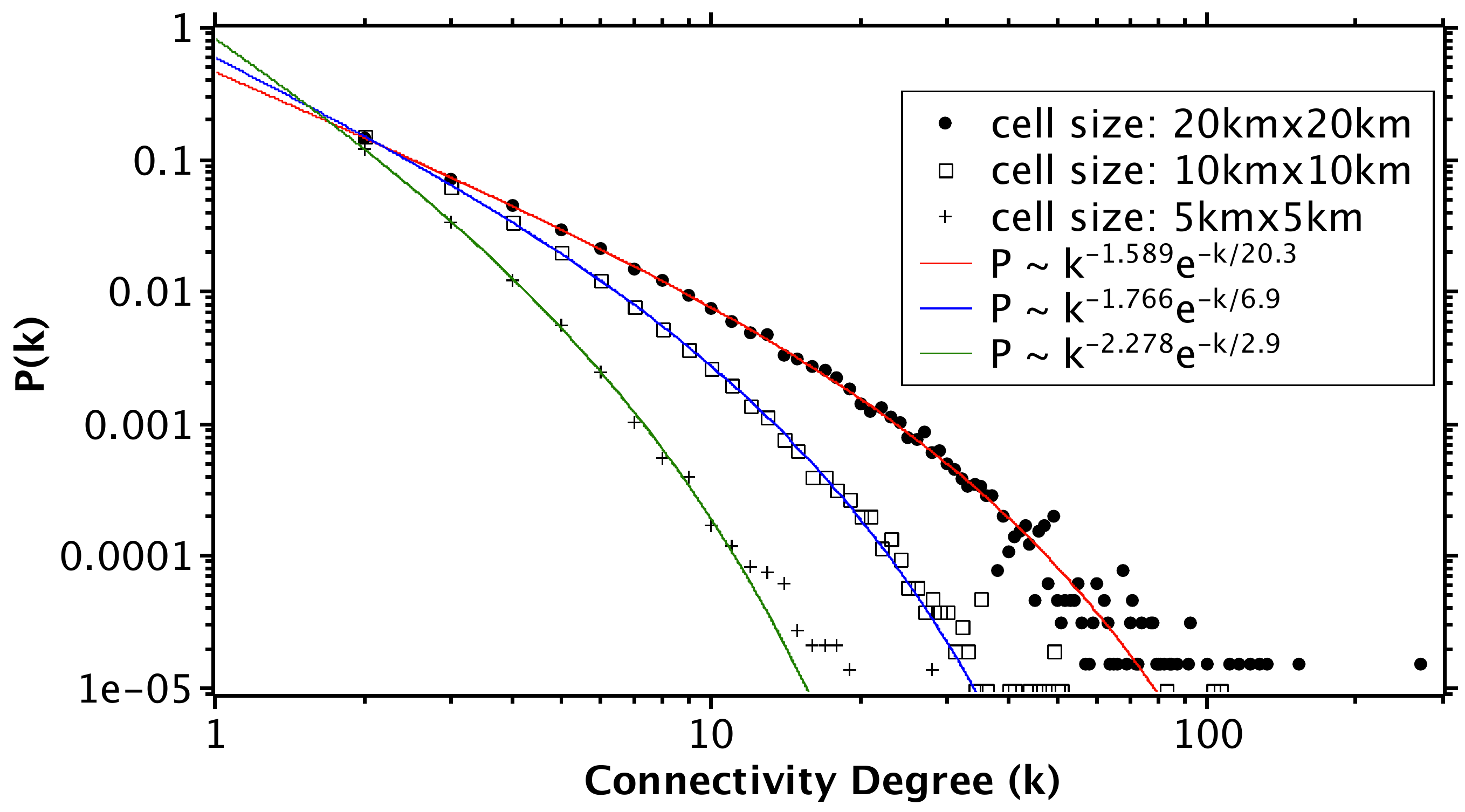}
      \label{k_globe_all}}
    \subfigure[]  {
    \includegraphics[width=0.6\columnwidth]{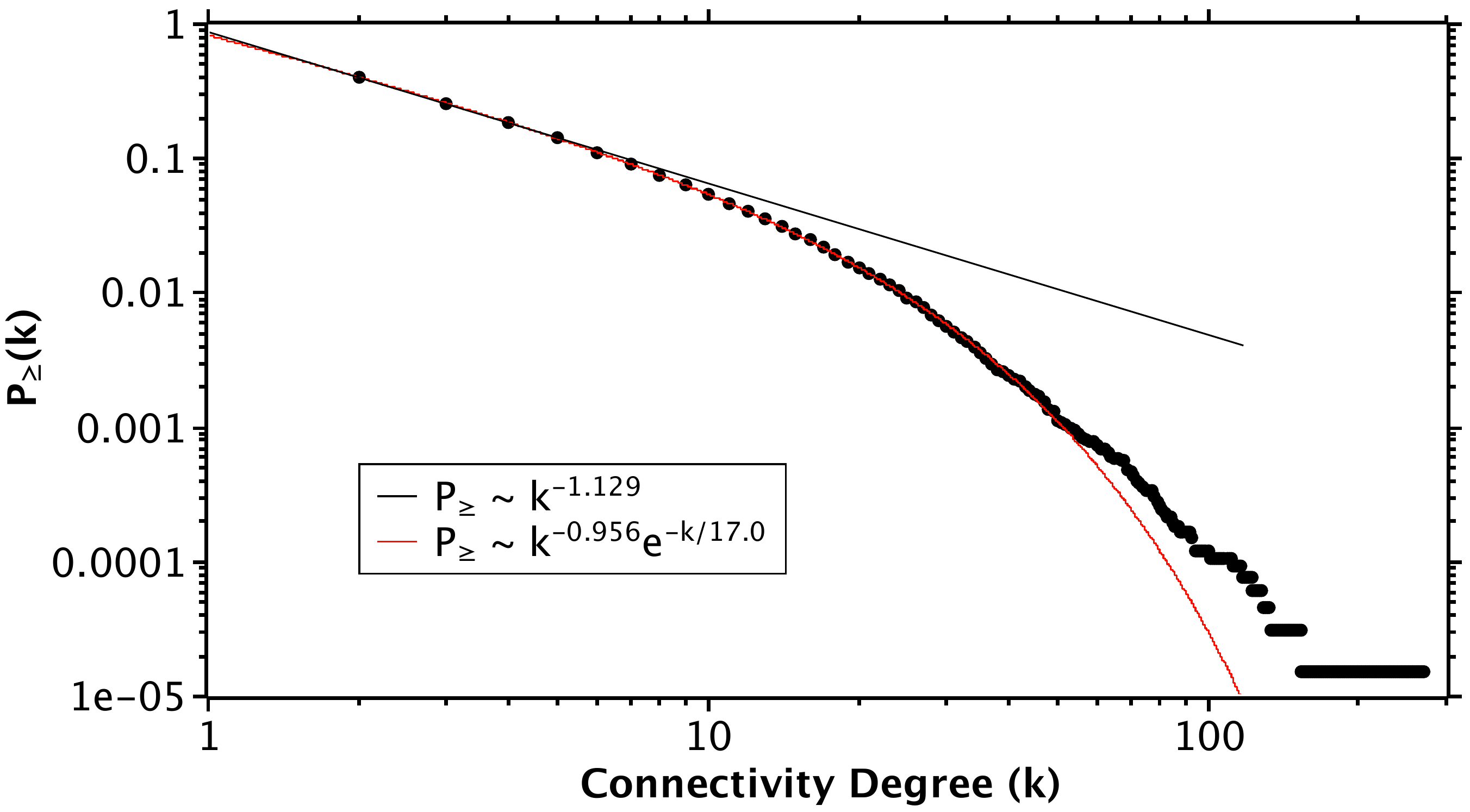}
      \label{k_globe_20}}
    \caption{Connectivity distributions in the global earthquake network. \subref{k_globe_all} Plot for the cell sizes $\text{20\,km} \times \text{20\,km}$ (solid circles), $\text{10\,km} \times \text{10\,km}$ (squares) and $\text{5\,km} \times \text{5\,km}$ (cross), where the solid lines represent the best fit using power-law with exponential cutoff. \subref{k_globe_20} Cumulative probability for the cell size $\text{20\,km} \times \text{20\,km}$. The solid lines represent a standard power-law (black) and a power-law with exponential cutoff (red). The best fit is the power-law with exponential cutoff with $\alpha = -0.956 \pm 0.001$ and $k_{c}= -17.0 \pm 0.1$. }
    \label{k_globe}
  \end{center}
\end{figure}

Looking at the world earthquake network constructed using the data described in Section~\ref{construction}, we note that the aging-cost effect are visibly stronger in the connectivity distribution; the exponential cutoff is clearly visible in both the degree distribution and the cumulative degree distribution presented in Fig.~\ref{k_globe}.

Fig.~\ref{k_globe_all} represents the connectivity distribution for the global networks using the three different cell sizes for the global lattice. It is interesting to note that, comparing these plots, we observe that the behavior is the same in all three cases (in the sense that they present a power law with an exponential cutoff), which indicates that the cell size does not change the complex features behind the global seismic phenomena. 

In Fig.~\ref{k_globe_20}, we have the same plot of
Fig.~\ref{k_globe_all}, but using the cumulative probability only for
cell sizes of $\text{20\,km} \times \text{20\,km}$. Note that the
cumulative probability plot for the global network shows the same
exponential cutoff behavior than for local network, as shown in
Fig.~\ref{k_california}.

\begin{figure}[h]
  \begin{center}
    \subfigure[] {
    \includegraphics[width=0.6\columnwidth]{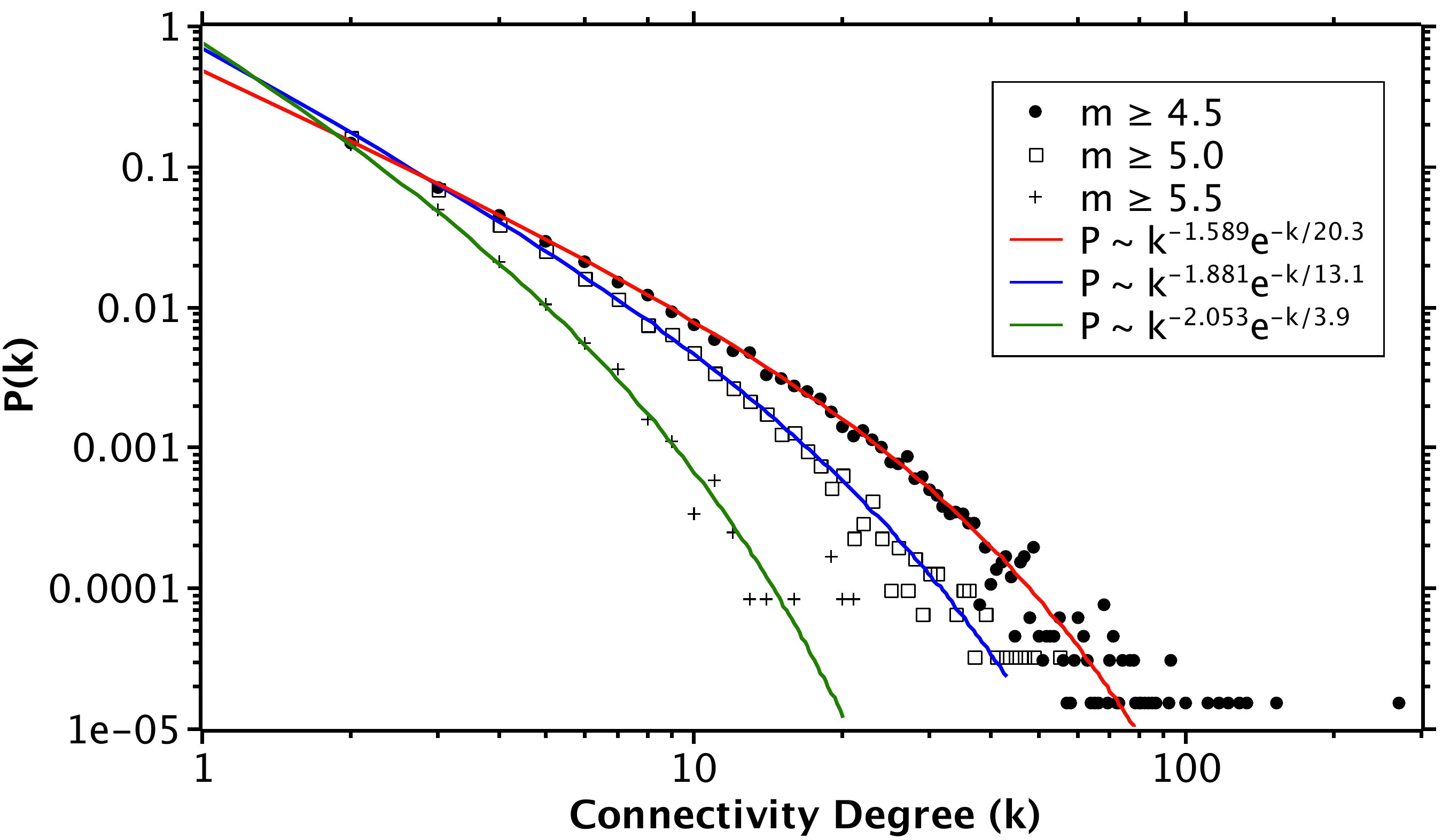}
      \label{freq_globe_magmin}}
    \subfigure[]  {
    \includegraphics[width=0.6\columnwidth]{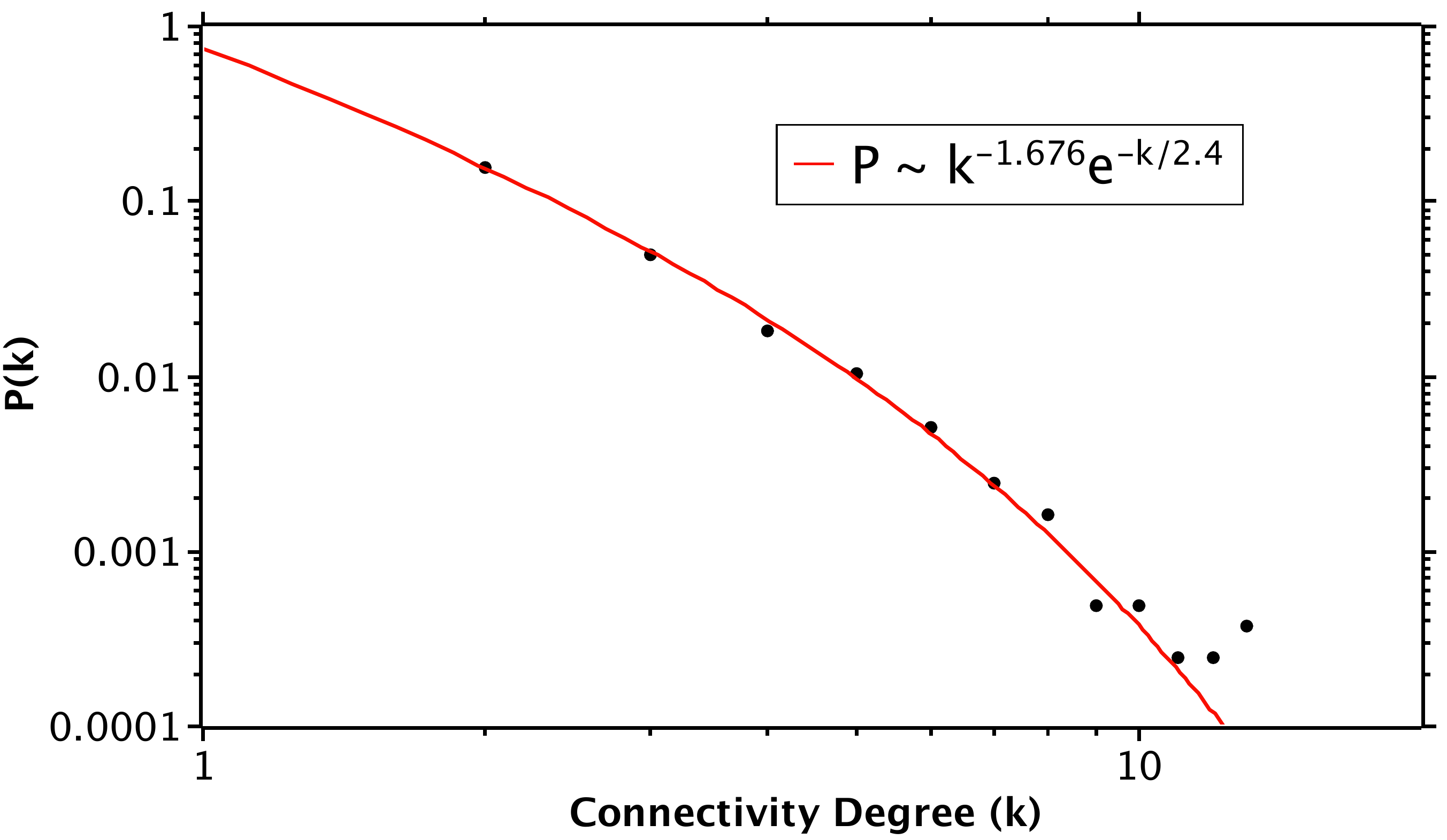}
      \label{freq_globe_1987}}
    \caption{Connectivity distributions in the global earthquake network using cell size $\text{20\,km} \times \text{20\,km}$. \subref{freq_globe_magmin} Plot for  $m \geq 4.5$ (solid circles), $m \geq 5.0$ (squares) and $m \geq 5.5$ (cross), where the data spans the time interval between 1972 and 2011. The solid lines represent the best fit using power-law with exponential cutoff. \subref{freq_globe_1987} Plot for $m \geq 5.5$ using data between the years 1987 and 2011. The solid line represents a power-law with exponential cutoff. }
    \label{k_globe_magmin}
  \end{center}
\end{figure}

It is noteworthy that, in order to show the consistence of our results, we have made two tests in our global network of epicenters. Firstly, to verify if the value that we considered as lower threshold of magnitude (4.5, in Richter scale) is satisfactory, we did the same analysis in the connectivity distributions using different magnitude thresholds for the globe. The magnitude intervals considered were $m \geq 4.5$, $m \geq 5.0$ and $m \geq 5.5$, where the number of nodes in the network in each case are 65\,355, 30\,763 and 11\,887, respectively. As we can see in Fig.~\ref{freq_globe_magmin}, in all cases the distributions have presented behaviors similar to those shown in Fig.~\ref{k_globe}, i.e, a power-law with exponential cutoff. Secondly, we check whether the technological deficiencies in the 70's and early 80's with respect to the detection of earthquakes have a relevant influence on our results. To do this, we plot the connectivity distribution using only seismic data between 1987 and 2011, for the magnitude interval $m \geq 5.5$ (the number of nodes in that network is 8\,112). Observing Fig.~\ref{freq_globe_1987} it is possible to note that we still have a power-law with exponential cutoff behavior.


\subsection{Small-world Property of the Global Seismic Network}
\label{small_world}

Small-world networks \citep{watts1999networks} have the general
characteristic that they contain groups of near-cliques (dense areas
of connectivity) but long jumps between these areas
(i.e. bridges). These two properties lead to a network in which the
{\it{average shortest path}} is very small and the {\it{clustering
    coefficient}} very high. It is important to note that the term
{\it{average shortest path}} does not refer to a spatial distance but
the number of ``steps'' on the network to move from a node to another.

Here we would like to test if the global seismic network has small-world properties. The consequence of such a finding would be an indicative that seismic events around the world are correlated and not independent. To study these properties we need to introduce slight changes to our original network. The first is that the loops have to be removed, since we are looking for correlations between nodes and it only makes sense when these nodes are different. The second change is, that we move from a network with multi-graph characteristics to a weighted network. That is, if two nodes are linked by $w$ edges in the original network, they will be linked by a single edge with weight $w$ in the new version of the network.

We have analyzed the seismic network for the entire world under two viewpoints: directed and undirected. The cell size used in this construction was $\text{20\,km} \times \text{20\,km}$. The data used was the same as described in Section~\ref{construction}. Table~\ref{table_small} shows the results obtained for the clustering coefficient ($C$) \citep{barrat04} and the average path length ($\ell$) \citep{brandes01}.


\begin{table}[h]
  \caption{Results for the clustering coefficient ($C$) compared to the clustering coefficient of a random network of the same size ($C_{\text{rand}}$) and the average path length ($\ell$) compared to the $\ln N$, where $N$ is 65\,355 in network with cell size $\text{20\,km} \times \text{20\,km}$.}
  \begin{center}
    \resizebox{0.6\columnwidth}{!}{%
      \tabcolsep=0.2cm \begin{tabular}{l c c c c}  
        \hline
        Network 	&  $C$ & $C_{\text{rand}}$ & $\ell$ & ln$N$	\\
        \hline
        Directed &  $\text{7.0}\times \text{10}^{\text{-3}}$ & $\text{4.2}\times \text{10}^{\text{-5}}$ & 17.19 & 11.08  \\
        Undirected	 & $\text{4.2}\times \text{10}^{\text{-2}}$ & $\text{4.2}\times \text{10}^{\text{-5}}$ & 12.24 & 11.08  \\
        \hline
      \end{tabular}%
    } 
  \end{center} 
  \label{table_small} 
\end{table}

From Table~\ref{table_small}, we note that both versions of the earthquake network have small-world properties; the clustering coefficient is much higher than an equivalent for a random network, and the average path length has the same order of magnitude as the logarithm of the number of nodes. It is worth noticing that the regional earthquake networks built for California, Japan and Chile also are small-world \citep{abe06,abe10} although the significance of small world at the global level is higher because with these worldwide results we have an indicative of long-range relations between different places around the world.

\subsection{Time interval between successive seismic events}
\label{sec:time_interval}

We have also studied the relationship between different regions of the world under another viewpoint, which is based on the analysis of the time intervals between the successive earthquakes in our global network.

Previous studies have found that the probability distribution of time intervals between successive seismic events in small areas of the world (e.g. California and Japan) can be well described by nonextensive statistical mechanics \citep{abe03,abe05,papadakis2013evidence,michas2013non,vallianatos2013evidenceSantorini,vallianatos2012non}. We will verify in this section if these features are still present when we look to the entire world.

In \citep{abe05}, the authors use concepts from nonextensive statistical mechanics to show that the cumulative probability distribution, $P_{\geq}$, for the time interval between successive earthquakes in California and Japan follows a $q$-exponential,
\begin{equation}
\label{prob_time_cum_distrib}
P_{\geq}( \Delta t) = e_{q}(-\beta \Delta t ),
\end{equation}
where, $\Delta t$ is the time interval between successive events and $\beta$ is a positive value. 

From Eqs. \ref{q_exp} and \ref{prob_time_cum_distrib} , it is possible to see that, if $q > 1$, when $\Delta t \gg [\beta (1 - q)]^{-1}$, the cumulative distribution represented in Eq.~\ref{prob_time_cum_distrib} approaches a power-law given by, $P_{\geq}( \Delta t) \sim \Delta t^{1/(1-q)}$.

Furthermore, from Eq. \ref{q_ln}, if we take the $ln_{q}$ in both sides in Eq.~\ref{prob_time_cum_distrib},
\begin{equation}
\label{q_ln_prob}
ln_{q}(P_{\geq}( \Delta t)) = - \beta \Delta t,
\end{equation}
we can observe that the $q$-logarithmic of $P_{\geq}( \Delta t)$ is linear with $\Delta t$ with a slope $-\beta$.

\begin{figure}[h]
\begin{center}
   \subfigure[]
       {\includegraphics[width=0.6\columnwidth]{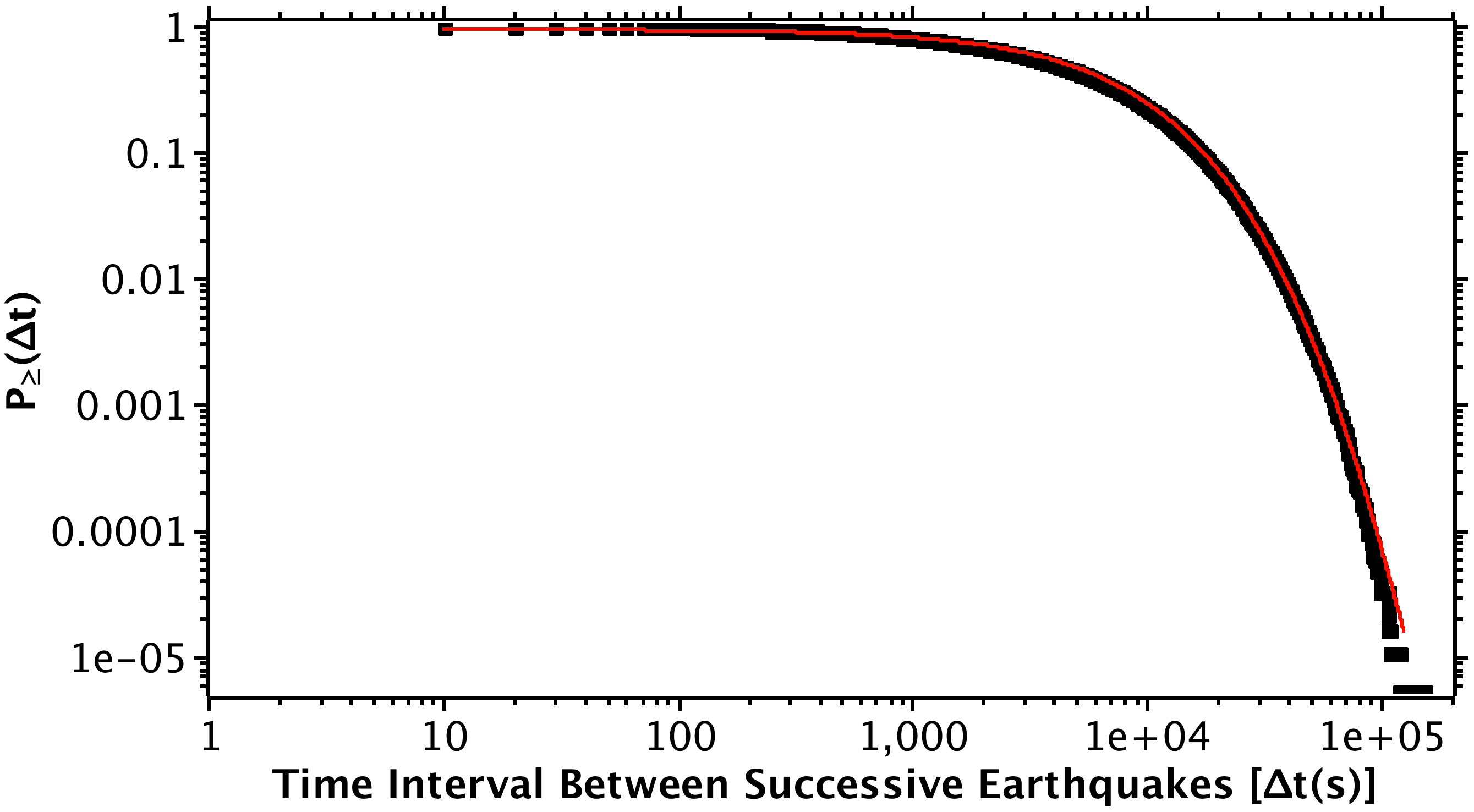}
       \label{log_time_interval}}\\
   \subfigure[]    
       {
\includegraphics[width=0.6\columnwidth]{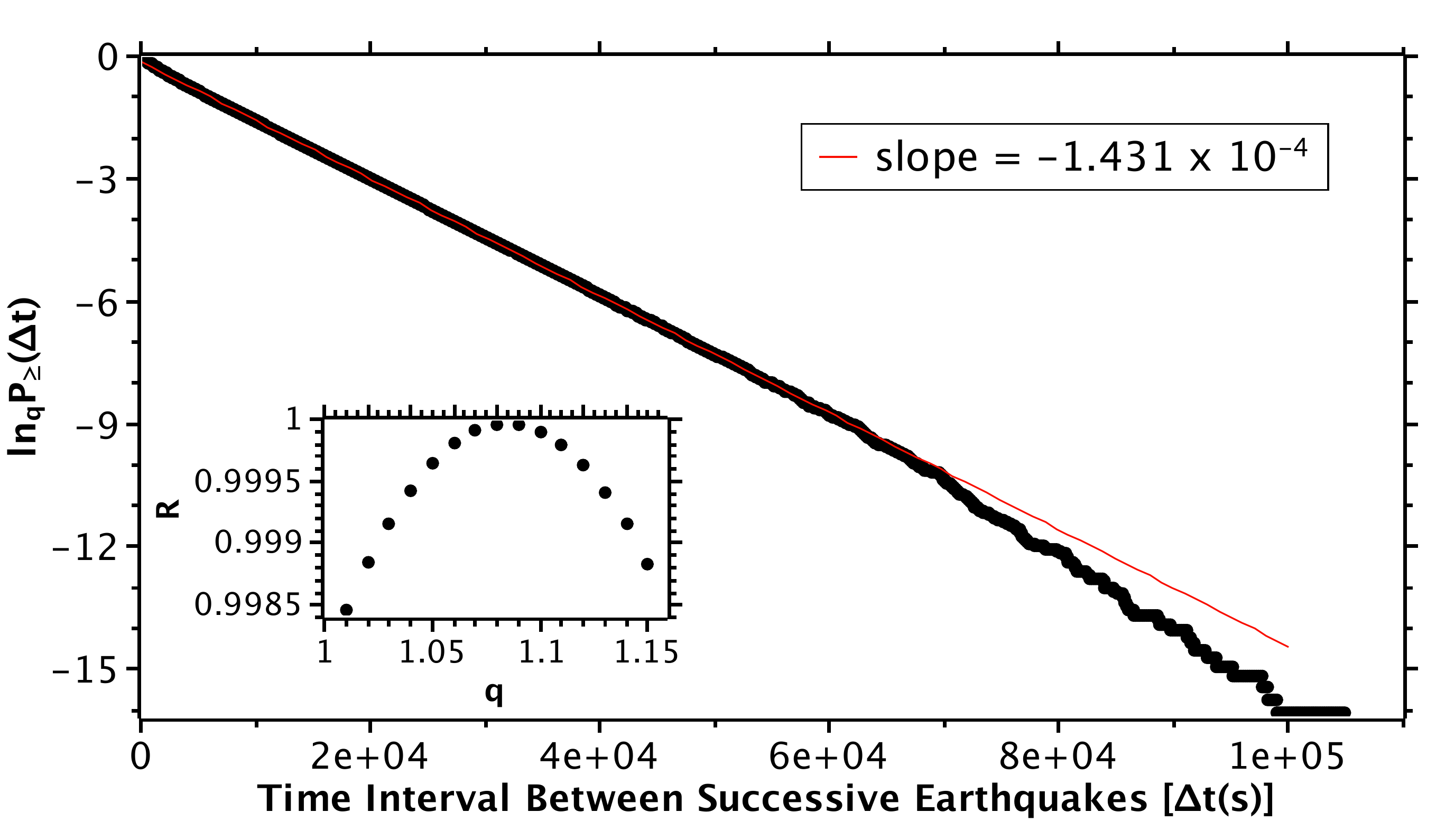}
       \label{qln_time_interval}}\hfill
\caption{Cumulative distribution of the time intervals between successive earthquakes with $m \geq 4.5$ in the entire world. \subref{log_time_interval} $log$-$log$ plot. The solid red line represents a $q$-exponential with $\beta = 1.431 \times 10^{-4}$ and  $q=1.08$. \subref{qln_time_interval} $q\,log$-$lin$ plot, where the black dots represent the data and the red straight line represents the best fitting using the Eq. \ref{q_ln_prob}. The slope of this plot gives $\beta = 1.431 \times 10^{-4} \pm 1 \times 10^{-7} s^{-1}$. Inset: linear correlation coefficient for some values of $q$. The best fit is obtained with $q=1.08$.}
\label{time_interval}
\end{center}
\end{figure}

Taking the worldwide data from Section~\ref{construction}, we plotted the cumulative probability distribution for the time interval between successive earthquakes and we noticed that this distribution is well fitted by a $q$-exponential, indicating that the nonextensive behavior is also present when we look at seismic events from a global perspective. Fig.~\ref{log_time_interval} shows the cumulative distribution on a $\log$-$\log$ plot, where the histogram was made by using bins of size equal to 10 seconds. In Fig.~\ref{qln_time_interval} we have the same distribution in a $q\,\log$-$\text{linear}$ plot, where the best value of $q$ was found by analyzing the values of the correlation coefficients, as shown in the inset of Fig.~\ref{qln_time_interval}. We remark here that, unlike our connectivity studies, which are not affected by the choice of a threshold (the number of connections of each cell in the subset under consideration is not affected by the occurrence of earthquakes below the threshold), our study on times between consecutive earthquakes could be affected by these occurrences. Through the threshold, we are favoring longer times over shorter time intervals. In this sense, the slope in Fig. 5b must be considered an upper limit for the actual value. A similar situation is observed in geomagnetic reversals where some short  chrons can be experimentally missed. In any case, this issue deserves an exclusive attention and the results of our research on it will appear elsewhere.

The results shown in this section are interesting because support the idea that there are connections between scale-free networks and non-extensive mechanics, as previously proposed \citep{thurner2005,soares2005}.

\section{Conclusions}  

The use of networks to model and study relationships between seismic
events has been used in the past for small areas of the globe. Here we
demonstrate that similar techniques could also be used at the global
level. More importantly, many of the techniques used in complex
network analysis were used here to show that there seem to exist
long-distance relations between seismic events which is a novel result
and not possible to be drawn from the previous studies for small areas
of the globe.

We have argued in favor of the long-distance relation hypothesis by
showing that the network has small-world characteristics. Given the
small-world characteristics of high clustering and low average path
length, we were able to argue that seisms around the world appear not
to be independent of each other. To strengthen this argument, we
decided to do a temporal analysis of our network. Plotting the
probability distribution for the time intervals between successive
earthquakes, we have found that this distribution is well fitted by a
$q$-exponential, indicating a behavior described by the non-extensive
statistical mechanics, which obtain $q$-exponential distributions from
the generalized Tsallis entropy. This non-extensive behavior also
contributes to the long-distance relation hypothesis, since the
non-extensive statistical has been used to explain many complex
systems with long-range interactions and long-range temporal
memory. Furthermore, our results contribute for the conjecture of the
connections between scale-free networks and non-extensive statistical
mechanics.

Another interesting approach we intend to do in the future relates to
using community analysis or community detection to understand how
seismic locations are grouped and the correlation of these groups with
active areas of seismic events around the world.

\section{Acknowledgements}

The authors are indebted to two anonymous referees whose comments and criticism have greatly contributed to improve the final presentation of this work. D.S.R.F thanks the Capes Foundation, Ministry of Education of Brazil, for the scholarship under process BEX 13748/12-2. A.R.R.P thanks CNPq (Brazilian Science Foundation) for his productivity fellowship.

\section{References}
\bibliographystyle{elsarticle-num} 
\bibliography{bib}


\end{document}